\documentclass[aps,prx,twocolumn,unsortedaddress,floatfix,nofootinbib,superscriptaddress]{revtex4-2}
\usepackage[mathlines]{lineno}
\usepackage{amsthm}
\usepackage{amsmath}
\usepackage{amssymb}
\usepackage{graphicx}
\usepackage{bm}
\usepackage{color}
\usepackage[dvipsnames]{xcolor}
\usepackage{enumitem}
\usepackage{mathrsfs}
\usepackage{verbatim}
\usepackage{bbold}
\usepackage{braket}
\usepackage{lipsum}
\usepackage[colorlinks, breaklinks, 
linkcolor=OrangeRed,
citecolor=RoyalBlue,
urlcolor=RoyalBlue]{hyperref}
\usepackage{ulem}

\usepackage{color} 

\newcommand{\blue}[2][]{\textcolor{blue}{#1#2}}

\usepackage{tikz}
\usepackage{natbib}
\usepackage{graphicx}

\newcommand{\tr}{\text{Tr}}
\newcommand{\ave}[1]{\langle #1 \rangle}

\newcommand{\sx}{\sigma^x}

\newcommand{\sz}{\sigma^z}

\begin{document}
\title{Reconstructing effective Hamiltonians \\
from nonequilibrium (pre-)thermal steady states}
\author{Sourav Nandy}
\affiliation{Jo\v{z}ef Stefan Institute, 1000 Ljubljana, Slovenia}
\author{Markus Schmitt}
\affiliation{Forschungszentrum J\"ulich GmbH, Peter Gr\"unberg Institute,
Quantum Control (PGI-8), 52425 J\"ulich, Germany}
\affiliation{University of Regensburg, 93053 Regensburg, Germany}
\author{Marin Bukov}
\affiliation{Max Planck Institute for the Physics of Complex Systems, N\"othnitzer Str. 38, 01187 Dresden, Germany}
\author{Zala Lenar\v{c}i\v{c}}
\affiliation{Jo\v{z}ef Stefan Institute, 1000 Ljubljana, Slovenia}
\date{\today}   

\begin{abstract} 
Reconstructing Hamiltonians from local measurements
is key to enabling reliable quantum simulation: both validating the implemented model, and identifying any left-over terms with sufficient precision is a problem of increasing importance.
Here we propose a deep-learning-assisted variational algorithm for Hamiltonian reconstruction by pre-processing a dataset that is diagnosed to contain thermal measurements of local operators. We demonstrate the efficient and precise reconstruction of local Hamiltonians, while long-range interacting Hamiltonians are reconstructed approximately.
Away from equilibrium, for periodically and random multipolar driven systems, we reconstruct the effective Hamiltonian widely used for Floquet engineering of metastable steady states. 
Moreover, our approach allows us to reconstruct an effective quasilocal Hamiltonian even in the heating regime beyond the validity of the prethermal plateau, where perturbative expansions fail.
\end{abstract}

\maketitle
\section{Introduction}
\label{intro}

A central idea at the core of modern quantum simulation is the utilization of highly controllable quantum degrees of freedom to emulate the behavior of complex quantum systems~\cite{Lewenstein2007, JAKSCH200552, Blatt2012, Schneider2012, Korenblit2012, Islam2013, Simon2011, Gambetta2017, BlochRMP2008, Jordens2008, Gross2017, Mazurenko2017, Zohar_2016}. Besides developing the corresponding state preparation and stabilization techniques, this requires the implementation of the Hamiltonian 
that governs the physics of the system 
to be simulated, in the first place. 
To do this, different approaches have been developed, which leverage specific properties of the underlying quantum system. For instance, analog simulators, such as ultracold atoms \cite{BlochRMP2008, Bloch2012, Aidelsburger2013, BLOCH20051, Gross2017} and trapped ions \cite{Blatt2012, Korenblit2012, Schneider2012}, are natural platforms of short- and long-range interacting Hubbard and Ising Hamiltonians. By contrast, the digital approach to quantum simulation \cite{Feynman1982-FEYSPW, Llyod} implemented in superconducting qubits, Rydberg atom \cite{Heras2014, Barends2015, Lamata2018, Weimer2011, Muller2012}  or trapped ion \cite{doi:10.1126/science.1208001, Muller2012, Barreiro2011} platforms, relies on constructing a time-evolution operator using consecutively applied discrete unitary gates.

Whenever the desired Hamiltonian cannot be implemented using the available building blocks in a straightforward way, nonequilibrium techniques, such as Floquet engineering~\cite{goldman2014periodically,bukov2015universal,eckardt2017colloquium} or dynamical decoupling~\cite{viola1998dynamical,viola1999dynamical,bylander2011noise,choi2020robust}, become handy. Recently, periodic drives have enabled the realization of non-equilibrium phase transitions~\cite{EckardtPRL2005, ZenesiniPRL2009}, topologically non-trivial systems~\cite{OkaPRB2009, KitagawaPRB2010, Jotzu2014, Lindner2011,sun2023engineering}, artificial gauge fields~\cite{AidelsburgerPRL2011,StruckPRL2012,BermudezPRL2011} and discrete time crystals~\cite{Time_crystal_Nayak_rev,Khemani2019, Sacha_2018, Mi2022, Zhang2017, Choi2017, Lesanovsky2019,beatrez2023critical}.  
Note that having an effective Hamiltonian is not guaranteed for generic nonequilibrium drives due to the lack of energy conservation; the necessary ingredients for the existence of effective Hamiltonians are currently the subject of intensive on-going research.  

\begin{figure}[t!]
\includegraphics[width=\columnwidth]{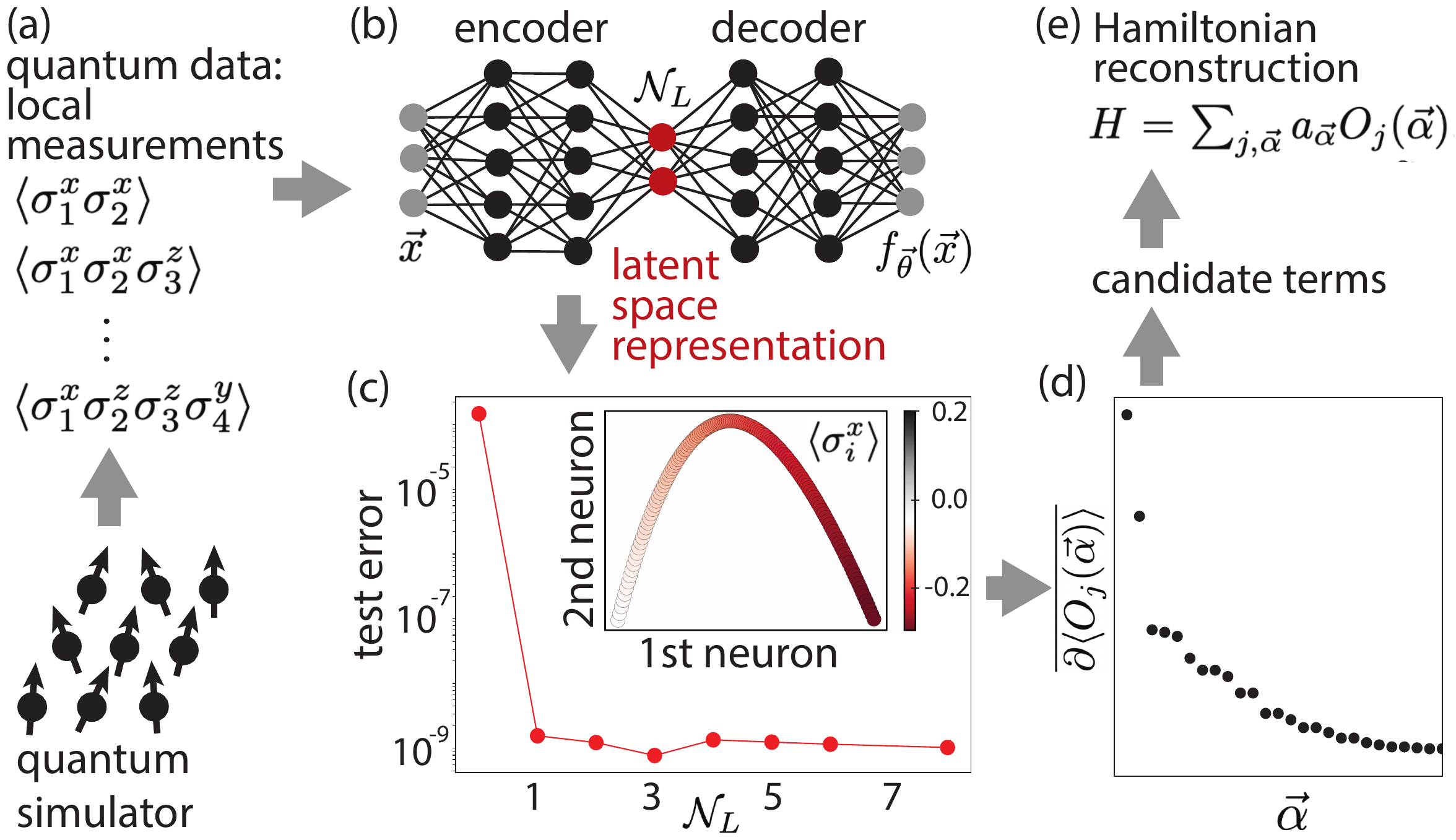}
\caption{Schematic presentation of the Hamiltonian reconstruction algorithm proposed in this work. (a) The input for the Hamiltonian reconstruction are measurements of local operators performed in (locally) (pre-)thermal (steady) states at different temperatures, for example, obtained from a quantum simulator. (b) These serve as input data for the autoencoder, which (c) confirms that the data is indeed parametrized with a single parameter (temperature) by checking that the decoding `test error' drops at a single neuron in the latent space, and (c, inset) to generate the latent representation of the dataset, which is one-dimensional for thermal data. (d)~In the next step, the potential candidate terms $O_j(\vec\alpha)$ in the Hamiltonian 
$H=\sum_{j, \vec\alpha}a_{\vec\alpha}O_{j}(\vec\alpha)$
are pre-selected as the operators with the largest average gradient of expectation values $\overline{\partial \ave{O_j(\vec\alpha)}}$ along the latent representation.
(e) In the last step, the Hamiltonian coefficients $a_{\vec\alpha}$ are fixed.
Here, we show actual data from effective Hamiltonian reconstruction in the prethermal plateau for our Floquet protocol, Eq.~\eqref{Prethermal_protocol}, presented in Fig.~\ref{fig:BCH_compare}.
}
\label{fig:schematic}
\end{figure}

Whichever the implementation method may be, engineered effective Hamiltonians often come with additional, unwanted terms. These can be residual (longer-range) interactions, or leftover single-particle field terms; they can also arise due to coupling to existing higher-energy levels or additional cavity modes \cite{johansen22}.
Moreover, nonequilibrium protocols used for Hamiltonian engineering rely on the existence of prethermal (meta-)stable steady states, in which heating (i.e., energy absorption and entanglement production) is suppressed up to controllably long times~\cite{MoriPRL2016,KUWAHARA201696,AbaninPRB2017_heating,MarinPRL2015,morningstar2022universality}. Nevertheless, residual terms in the effective Hamiltonian (e.g., higher-order corrections) eventually lead to accumulating detrimental heating.  
In parallel, verifying that a desired Hamiltonian is implemented to a given precision, is a crucial requirement for a reliable quantum simulation; moreover, certification of the realized Hamiltonians becomes indispensable for benchmarking the performance of present-day noisy intermediate scale quantum (NISQ) devices~\cite{AltmanPRX2021, Bloch2012, Daley2022, Brigel2008, Guerreschi_2020}.  
For all these reasons, developing a procedure, based on measurements of local observables, and capable of identifying and quantifying the Hamiltonian implemented in quantum simulators to a high precision, is a problem of ever-increasing importance. 

During the last few years, several methods for Hamiltonian reconstruction have been put forward: from measurements on a single eigenstate~\cite{Qi2019, ChertkovPRX2018, GreiterPRB2018, LindlerPRL2019} using trusted quantum simulator~\cite{Granade_2012, WiebePRL2014, Wang2017}; or from eigenstate dynamics~\cite{NoriPRA2009,FrancoPRL2009,SarovarPRL2014,Kim2016,SonePRA2017,IEEE, Li2020}, via compressed sensing~\cite{RudingerPRA2015}, using restricted Boltzmann machine tomography~\cite{WiebePRA2017} and neural networks~\cite{Valenti2019}. One common feature of these techniques is that they explicitly assume the underlying Hamiltonian to be both local and a translationally invariant sum of few-body terms. 
Furthermore, only a few of the above-mentioned methods are applicable to Floquet systems. Although reasonable progress has been achieved recently in this direction \cite{PRXQuantum.3.030324,Olsacher_2022}, the methods employed in these works demand an apriori intuition from the Floquet-Magnus expansion for the selection of candidate terms. However, a typical challenge in any Hamiltonian reconstruction algorithm often lies in developing a systematic method to select candidate terms. It is, therefore, highly desirable to propose an algorithm that relies on an unbiased approach to select the candidate terms of the target Hamiltonian.



In this work, we propose a machine learning-assisted algorithm for the reconstruction of Hamiltonians, which uses as the input the expectation values of local observables that are diagnosed to be thermal. 
We show that our algorithm can precisely reconstruct local Hamiltonians, while long-ranged Hamiltonians are approximately reproduced with certain limitations.
It is worth mentioning that the presence of only few-body terms in the underlying Hamiltonian is advantageous for our method but it is not a necessary requirement.
We apply our algorithm to Floquet and random multipolar driven systems with the purpose of reconstructing the effective Hamiltonian, responsible for the stroboscopic expectation values in the prethermal plateau and/or in the follow-up heating regime. For constructing the effective Hamiltonian in the prethermal plateau, our algorithm does not require any a priori intuition from the Floquet-Magnus expansion or any other similar perturbative technique when the selection of candidate terms is concerned. 
Furthermore, we go beyond prethermal steady states and analyze the heating regime, where perturbative expansions are known to fail. We extract an effective, quasistatic Hamiltonian, which reproduces the thermal measurements of local operators as the system approaches the infinite-temperature state. We directly observe that this effective Hamiltonian becomes less and less local as the system heats up, verifying a hypothesis laid out in previous work~\cite{MoriPRL2016, bukov2016heating}. 

The rest of the paper is organized as follows. In Sec.~\ref{secII}, we discuss in detail the algorithm of Hamiltonian reconstruction. Sec.~\ref{secIII} deals with illustrative examples where our method is used to reconstruct both short and long-range interacting static Hamiltonian. This is followed by the central Sec.~\ref{secIV}, which addresses Hamiltonian reconstruction in dynamical Floquet setups, starting with reconstruction in the prethermal regime. Thereafter, we discuss the insight our algorithm provides into the physics of unconstrained heating for both Floquet systems and non-Floquet systems that exhibit eventual thermal death. Finally, we conclude and present future perspectives in the Sec.~\ref{conclude}.

\section{Algorithm for Hamiltonian reconstruction using autoencoders}
\label{secII}

In our Hamiltonian reconstruction algorithm, we rely on a deep autoencoder neural network~\cite{Bourlard2004AutoassociationBM,  Tschannen2018}, Fig.~\ref{fig:schematic}(b), which is used for the dimensional reduction of the data; in our algorithm this amounts to a preselection of the Hamiltonian candidate terms. In our case, each element of dataset contains measurements of different local operators in a given quantum state, see Fig.~\ref{fig:schematic}(a). Our approach is applicable if these states are (effectively) thermal. The autoencoder is used to generate a latent representation of the dataset and to verify that the dataset indeed contains measurements on thermal states; whenever this is the case, the latent representation is a one-dimensional manifold, Fig.~\ref{fig:schematic}(c). Such compressed representation of measurement helps us single out the candidates terms of the Hamiltonian, Fig.~\ref{fig:schematic}(d), significantly reducing the complexity of the Hamiltonian reconstruction, Fig.~\ref{fig:schematic}(e), performed in the subsequent step. 
Previously, some of us proposed part of the algorithm as a way to reconstruct spatially local Hamiltonians \cite{LenarcicMLPRB2022}. Here we extend the procedure to deal with long-range Hamiltonians as well.
In the following, we explain the algorithm in more details. 


The autoencoder is a deep neural network, where the input and output layers have the same dimension, while at least one of the middle layers has a considerably smaller number of neurons $\mathcal{N}_L$ and represents a bottleneck, Fig.~\ref{fig:schematic}(b). The part prior to the bottleneck is called the encoder, which maps the input data to a lower-dimensional latent representation. The decoder then maps the latent representation to the output. 
The network is set by minimizing the difference between the input $\vec x$ and the output $f_{\vec\theta}({\vec x})$ with respect to the network parameters $\vec\theta$, averaged over a minibatch of data elements from the dataset $\mathcal D_T$, 
\begin{equation}
    \mathcal L_{\mathcal D_T}(\vec\theta)=
    \frac{1}{|\mathcal D_T|}\sum_{\vec x\in\mathcal D_T}
    \big(f_{\vec\theta}(\vec{x})-\vec{x}\big)^2.
    \label{eq:reconstruction_loss}
\end{equation}
The latent representation's dimension, i.e., the intrinsic dimension of the data, is extracted from the same function (called `test error'), evaluated for the optimized neural network parameters $\vec\theta$ on the rest of the unseen data elements. The intrinsic dimension of data is identified as the number of neurons in the bottleneck at which the `test error' drops significantly and is flat upon further increasing 
$\mathcal{N}_L$, Fig.~\ref{fig:schematic}(c). 
Values at neurons in the bottleneck give the latent representation of data. In the inset of Fig.~\ref{fig:schematic}(c), an example of latent represetation is shown for a network with $\mathcal{N}_L=2$ neurons in the bottleneck. 
Further technical details on our network are given in Appendix~\ref{App:Autoencodes}.

Our Hamiltonian reconstruction algorithm relies on precondition (0) and consists of Steps (1-5):
\begin{itemize}
\item[(0)] Work under the assumption that we are given thermal expectation values of all local operators with support $\ell$ that is smaller than or equal to some pre-set number $\mathcal{S}$. These operators are measured in states of different temperatures: the set of all measurements for a given state represents one data element, while different states yield the dataset.
\item[(1)] Use autoencoder to check that the data elements indeed contain measurements with respect to (effectively) thermal states. In that case, the latent representation of the dataset is one-dimensional, with data elements within it ordered with respect to the temperature, or equivalently, the energy of the corresponding states. 
\item[(2)] Since data elements in the latent representation are ordered with respect to the Hamiltonian expectation value, other operators that correspond to the individual terms in the Hamiltonian also show a pronounced variation along that one-dimensional manifold. Therefore, one can isolate candidate Hamiltonian terms $O_j(\vec\alpha)$, $H=\sum_{j, \vec\alpha}a_{\vec\alpha}O_{j}(\vec\alpha)$, via singling out those operators $O_{j}(\vec\alpha)=\sigma_{j}^{\alpha_{j}}...\sigma_{j-1+|\ell|}^{\alpha_{j-1+|\ell|}}$ that have the largest average gradient $\overline{\partial \ave{O_j(\vec\alpha)}}$ in the expectation value along the latent representation. We will assume periodic boundary conditions and a translationally invariant system with $O(\vec\alpha) \equiv O_{j=1}(\vec\alpha)$, but the procedure can in principle be performed in the absence of that as well.
\item[(3)] Step (2) will, in addition to the actual terms, filter out also some spurious ones, which are not part of $H$ but are, for example, products of actual Hamiltonian terms on neighboring sites. 
To fix the prefactors $a_{\vec\alpha}$, and to get rid of the spurious terms, we compare the `trial' thermal expectation values (with respect to the trial Hamiltonian) to the actual measurements for one data element at a given (but unknown) temperature $1/\beta$, and find the solution of the following equation 
\begin{equation}
\tr\left[O(\vec\alpha') \frac{e^{- \beta\sum_{j, \vec\alpha} a_{\vec\alpha} O_{j}({\vec\alpha})}}{Z}\right]
- \ave{O(\vec\alpha')}_{\beta} = 0.
\end{equation}
Since one does not know the temperature of the state, $\beta$ remains undetermined; what is actually extracted is the product $\beta a_{\vec\alpha}$; $\beta$ can be determined separately, e.g., from dynamical measurements.
\end{itemize}

For strictly local Hamiltonians and measurements that include all Hamiltonian terms (i.e., the maximal support of measured operators $\mathcal{S}$ is larger or equal to the support of $H$), Steps (1-3)  will reconstruct $H$ (up to a prefactor) with high precision (see examples below), as already proposed and tested in Ref.~\cite{LenarcicMLPRB2022}. 
In the next steps, we extend the algorithm from  Ref.~\cite{LenarcicMLPRB2022} to be applicable to data from long-range Hamiltonians as well.
For long-range interacting Hamiltonians, some error in the reconstruction is inevitable, stemming from the fact that we have measurements of only the most local terms in the Hamiltonian and that we can at most aim at reconstructing a local approximation of a long-range Hamiltonian. In that case, Steps (1-3) might still generate finite prefactors for ``ghost" terms within the measured support. To remove those, we add here additional steps:
\begin{itemize}
\item[(4)] 
Compare the solutions of Step (3) at a few different $\beta$, i.e., for different data elements. Ghost terms' coefficients will have a large relative variance across solutions for different states; hence, we drop terms with 
\begin{equation}
\frac{\mathrm{Var}_{\beta}(a_{\vec\alpha}/a_{\vec\alpha_0})}{\mathbb{E}_{\beta}(a_{\vec\alpha}/a_{\vec\alpha_0})} > O(1).
\end{equation}
where $a_{\vec\alpha_0}$ is the coefficient at the dominant term.

\item[(5)] Repeat Step (3) for the Hamiltonian ansatz without the ghost terms. This step just slightly corrects the result of Step (3).
\end{itemize}
The most expensive step of our reconstruction is Step (3); even after performing Step (2), potentially a large number of trial terms remain for Step (3). This can be particularly problematic for a Hamiltonian $H$ with large support. In order to ameliorate this issue to a manageable extent, one can use an iterative scheme:

First, perform Steps (2-4), selecting contributing operators $O(\vec\alpha)$ with maximal support $\mathcal{S}_k=\ell_k, k=0$: 
\begin{itemize}
\item[(k1)] When increasing the maximal support to $\mathcal{S}_{k+1}=\ell_{k+1}=\ell_k+1$, previously selected $\mathcal{S}_k$ support terms are retained by default, while Step (2) is performed only for operators with support $\ell_{k+1}$.
\item[(k2)] Steps (3-4) are performed, including previously selected $\mathcal{S}_k=\ell_k$ and new $\ell_{k+1}$ support trial terms from (k1). 
\item[(k3)] Terms surviving in (k2) are the new input for the next iteration.
\end{itemize}
In the end, Step (5) is performed.

\section{Reconstructing short- and long-range interacting static Hamiltonians}
\label{secIII}
We now benchmark our approach for two classes of $H$: local and long-range interacting Hamiltonians. 
The extent of applicability to the latter type of Hamiltonians has not been discussed in Ref.~\cite{LenarcicMLPRB2022}. 
In any case, each data element consists of measurements in a thermal state (see appendix \ref{Typicality} for details of how we prepare the initial thermal states) with temperature $1/\beta$ for all local operators with support smaller than or equal to $\mathcal{S}=4,6$. 
We use inverse temperatures from the interval $\beta\in [0.05,0.5]$.

First, we consider strictly local Hamiltonians with maximal support $\mathcal{R}$, for example,
\begin{eqnarray}
H_{L} = \sum_{i} \left( h_z \sigma^z_{i} + \sum_{r=1}^{\mathcal{R}-1}(J_1 \, \sigma^x_{i} \sigma^x_{i+r} + J_2 \, \sigma^y_{i} \sigma^y_{i+r})\right).
\end{eqnarray}
We find that the relative reconstruction error for coefficients $h_z, J_1, J_2$ highly depends on the maximal support of the measured operators $\mathcal{S}$: 
\begin{eqnarray}
& \mathcal{S} \ge \mathcal{R}: \text{relative error} \sim 10^{-6}, \\
& \mathcal{S} < \mathcal{R}: \text{relative error} \sim \mathcal{O}(1). \notag
\end{eqnarray}

Next, we consider long-range interacting Hamiltonians with power-law decaying interactions,
\begin{eqnarray}\label{eq:NL}
H_{LR}=\sum_{i} \left(
h_z \sigma^z_{i} + 
\sum_{r=1}^{\frac{N}{2}-1}\frac{J}{r^{\delta}}(\sigma^x_{i}\sigma^x_{i+r}+\sigma^y_{i}\sigma^y_{i+r})
\right)
\end{eqnarray}
with periodic boundary conditions.

\begin{figure}[t!]
\includegraphics[width=\columnwidth]{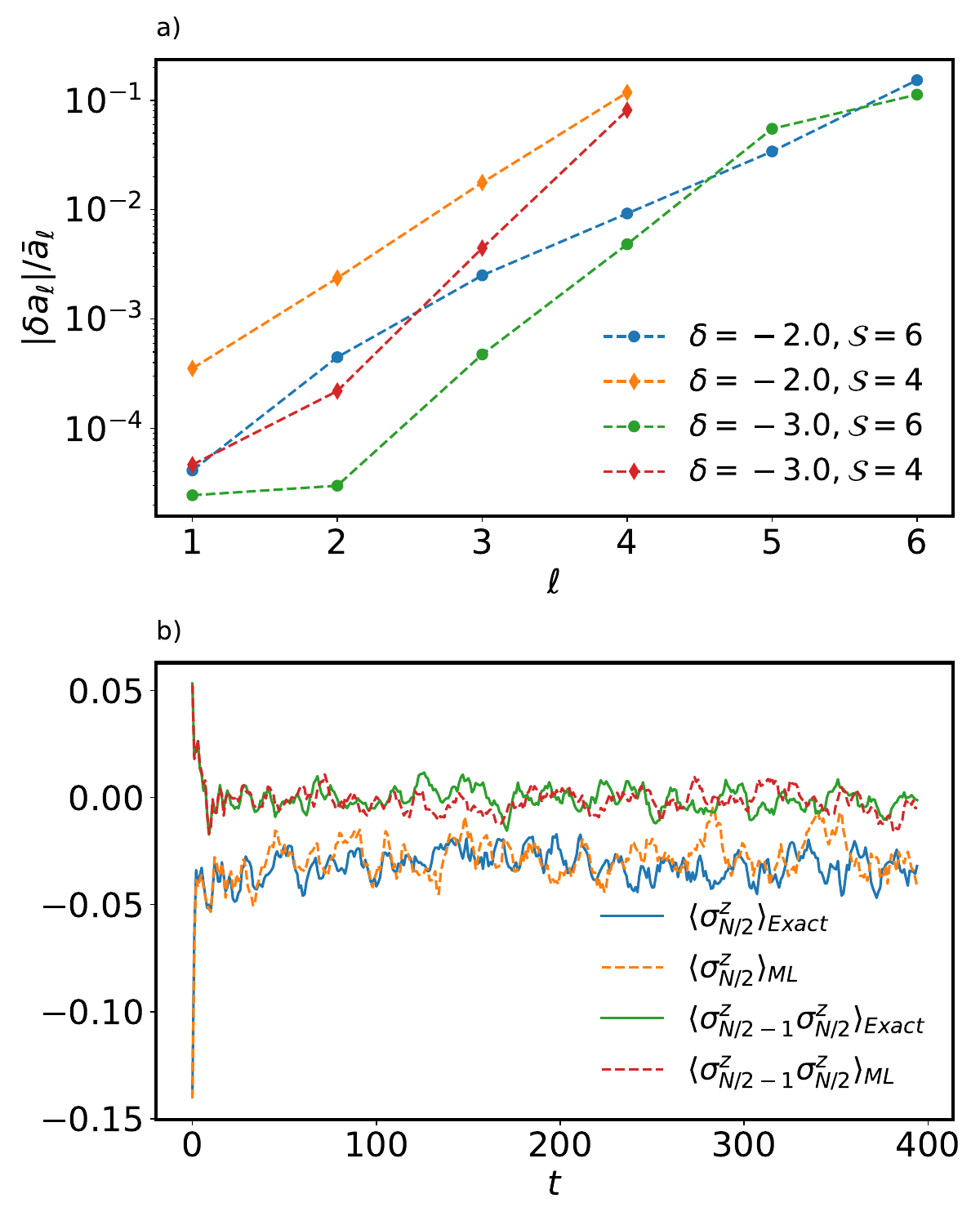}
\caption{
Evaluation of Hamiltonian reconstruction $H=\sum_{j, \vec\alpha}a_{\vec\alpha}O_{j}(\vec\alpha)$ for the long-range interacting Hamiltonian from Eq.~\eqref{eq:NL}. 
(a) The relative error ${|\delta a_\ell|}/{\bar{a}_\ell}$ of all reconstructed coefficients at a given support $\ell$. It is seen clearly that for a fixed $\delta$ (measure of long-rangedness of the Hamiltonian), terms with larger support have larger relative error. Also, for a given support $\ell$, relative error decreases with increasing $\delta$, i.e., considering more short-ranged underlying Hamiltonian.
(b) Time evolution of two different observables with respect to (i) the original long-range $H_{LR}$ with $\delta=3$ from Eq~\eqref{eq:NL}, and
(ii) its reconstruction $H_{\text{ML}}$ using our method. This figure clearly shows that even though Hamiltonian reconstruction is not perfect for long-ranged systems, the reconstructed Hamiltonian nonetheless captures local dynamics quite well.
Parameters: $N=16,  J=1.0, h_z=0.2, \delta=3.0$.
}
\label{fig:longrange}
\end{figure}

In this case, the reconstruction error highly depends on the maximal support of the measured operators $\mathcal{S}$ and the interaction strength decay exponent $\delta$. Fig.~\ref{fig:longrange}(a) shows the relative error ${|\delta a_\ell|}/{\bar{a}_\ell}$, for compact representation averaged over all reconstructed Hamiltonian terms with support $\ell$, i.e., 
\begin{align}
\label{relative error}
\frac{|\delta a_\ell|}{\bar{a}_\ell}&=\frac{|h_z-a_z|}{h_z}, \hspace{4mm} \ell=1 \\
\frac{|\delta a_\ell|}{\bar{a}_\ell}&= \frac{|2J/r^\delta-a_{xI\dots Ix} + a_{yI\dots Iy}|}{2J/r^\delta}, \hspace{4mm} \ell>1
\end{align}
We see that terms with larger support have larger relative error. In general, the latter increases for less local interactions (smaller $\delta$). However, even if the Hamiltonian is not exactly reconstructed, its effect at the level of dynamics is captured in practice. To support that, we show in Fig.~\ref{fig:longrange}(b) time evolution from a domain wall initial state for the original and the reconstructed Hamiltonian for a few local observables. For the sake of comparison, we use the value of $\beta$ that the data was measured at to divide the reconstructed coefficients $\beta a_{\vec\alpha}$ so as to obtain the actual Hamiltonian. In a temperature-agnostic protocol, $\beta$ would be determined as the one at which the comparison is best, as we demonstrate in the next section.

\section{Reconstruction of effective Hamiltonians engineered by nonequilibrium drives}
\label{secIV}
Having benchmarked our approach, we now turn to our central example: effective Hamiltonian reconstruction in dynamical systems, which is of interest for verification of quantum simulators, where desired Hamiltonians are often engineered using periodic protocols.
In these Floquet setups, the stroboscopic evolution is generated by the so-called Floquet Hamiltonian; while such closed systems are (in the absence of symmetries \cite{ArnabPRB2014}) bound to eventually heat up to a featureless infinite temperature state \cite{lazarides2014equilibrium,Rigol2014, Weidinger2017}, the heating process is delayed exponentially with increasing the frequency of the drive \cite{MoriPRL2016,AbaninPRB2017_heating}. When observed stroboscopically, at intermediate times, in the so-called prethermal plateau~\cite{ho2023quantum}, the system is locally effectively described by a thermal state with respect to an effective Hamiltonian that can feature 
nontrivial transient states \cite{MoriPRL2016,KUWAHARA201696,AbaninPRB2017_heating,MarinPRL2015}. Floquet protocols have thus been one of the central approaches in engineering non-equilibrium phenomena using quantum simulators and materials.

In the inverse frequency expansion or iterative rotating frame transformation, the effective Hamiltonian can be reconstructed up to some order \cite{Marin2015}. However, this is generally a daunting task \cite{Marin2015, KUWAHARA201696} and the former cannot be computed in closed form for generic systems and protocols. Here we propose an unbiased reconstruction, which does not rely on any expansion technique; instead, we reconstruct the effective Hamiltonian up to support $\mathcal{S}$, effectively resumming contributions from higher orders in the inverse frequency expansion within this support.
As the inverse frequency series  is formally divergent, the system eventually heats up; it is still not fully settled how this divergence is related to heating dynamics featured in the exact stroboscopic dynamics.
Our approach reveals that stroboscopically, throughout the whole heating regime, the system locally appears as if in a thermal state with respect to an effective quasi-static Hamiltonian; moreover, we demonstrate directly how its support grows in time.

\subsection{Floquet driving}
\begin{figure}[t!]
\includegraphics[width=\columnwidth]{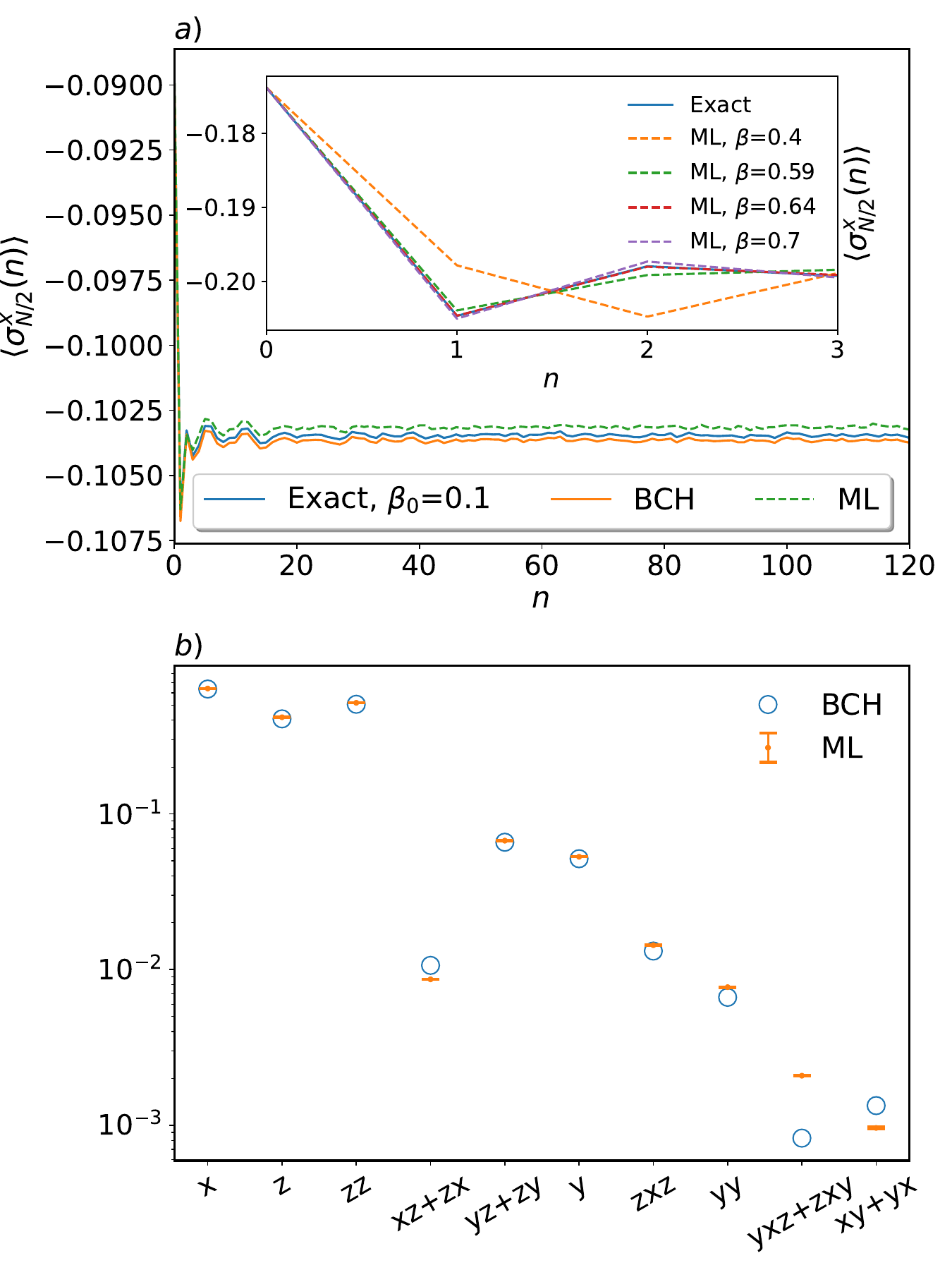}
\caption{(a) Stroboscopic measurements of $\ave{\sigma^x_{N/2}}$ as a function of cycle number $n$, obtained from: (i) the exact Floquet protocol, (ii) time evolution with respect to the BCH Hamiltonian up to second order in $T$, and (iii) time evolution with respect to our reconstruction $H_{\text{ML}}$. Initial states are thermal w.r.t.~$H_1$ at $\beta_0=0.1$. The inset demonstrates how we extract the overall multiplicative factor $\beta$ of the coefficients obtained from our Hamiltonian reconstruction algorithm, as discussed in the main text. 
(b) Comparison of Hamiltonian coefficients computed from
the BCH expansion and our Hamiltonian reconstruction algorithm. Clearly, our reconstruction algorithm predicts the coefficients (particularly the leading ones) quite accurately. Parameters: $J=1.0, h_x=0.9045, h_z=0.809, \gamma=0.4, \omega=2\pi/T=10.0, N=12$. 
}
\label{fig:BCH_compare}
\end{figure}

{\it Prethermal regime.}
In the following, we reconstruct the effective Hamiltonian, characterizing the stroboscopic pre-thermal state in the prethermal plateau, for the square pulse Floquet protocol
\begin{align}
\label{Prethermal_protocol}
U&= \exp\left(-iH_1T/2\right)\exp\left(-iH_2T/2\right) \\
\label{Pretherm_protocol_1}
H_1&=  \sum_{j=1}^{N} J\sigma^z_j \sigma^z_{j+1} + h_x \sigma^x_{j} + h_z \sigma^z_{j} \\ 
\label{Pretherm_protocol_2}
H_2&=\gamma \sum_{j=1}^{N}  \sigma^x_j
\end{align}
where $U$ is the unitary matrix for one drive cycle which acts repeatedly on the system during the dynamics; $T$ denotes the time period of the drive.
Fig.~\ref{fig:BCH_compare}(a) shows time-dependent expectation values $\langle\sigma^x\rangle$ for $J=1.0, h_x=0.9045, h_z=0.809, \gamma=0.4, \omega=2\pi/T=10.0$, for a time evolution from an initial thermal state with respect to $H_1$ and inverse temperature $\beta_0=0.1$.

For drive cycle numbers $n\gtrsim 5$, we observe an extended prethermal plateau. Visible oscillations are a consequence of finite size ($N=12$) exact diagonalization propagation.
In the large frequency regime $\omega \gg 1$ considered, the Floquet Hamiltonian can be perturbatively approximated using Baker-Campbell-Hausdorff (BCH) series expansion, as detailed in Appendix \ref{BCH}.

Our Hamiltonian reconstruction is performed on a dataset with elements containing measurement of local operators with maximal support $\mathcal{S}=3$, in states that are evolved up to some fixed time within the prethermal plateau. 
As initial states for the time evolution we consider thermal states with respect to $H_1$, with inverse temperatures within the range $\beta_0 \in [0.05,0.40]$.
In principle, one could start from non-thermal states as well since these are expected to relax to thermal states in the plateau anyhow, due to the nonintegrability of the exact Floquet Hamiltonian. 
While Step (3) in the reconstruction algorithm [cf.~Sec.~\ref{secII}] can in principle be performed for a single data element (a single state), we average the relative coefficients $a_{\vec\alpha}/a_x$ ($a_x$ being the dominant one) for each $\vec\alpha$ over a few $\beta_0$ and a few measurement times in the prethermal plateau.
As discussed in Sec.~\ref{secII}, our algorithm gives the coefficients up to an overall multiplicative factor $\beta$. This overall multiplicative constant $\beta$ can be extracted from dynamics in the initial time window before the system reaches the prethermal plateau; as illustrated in the inset of Fig.~\ref{fig:BCH_compare}(a). 
More precisely, we determine $\beta$ by minimizing the difference $\sum_{n} |\tr[O U^n \rho(0) (U^\dagger)^{n}] - \tr[O e^{-inTH_{\text{ML}}(\beta)} \rho(0) e^{inTH_{\text{ML}}(\beta)}]|$ between the exact and the reconstruction $H_{\text{ML}}$ based time-dependent measurements of a dominant operator at early cycles.

In Fig.~\ref{fig:BCH_compare}, our Hamiltonian reconstruction $H_{\text{ML}}$ is compared to a second order (in $T$) BCH expansion, which contains terms with the same maximal support $\mathcal{S}=3$.
Fig.~\ref{fig:BCH_compare}(a) compares the dynamics generated from (i) the exact unitary, Eq.~\eqref{Prethermal_protocol}, (ii) the Hamiltonian constructed from the BCH approximation calculated up to second order in $T$, as discussed in the Appendix \ref{BCH}, and (iii) our reconstructed Hamiltonian $H_{\text{ML}}$, for an initial state with $\beta_{0}=0.1$.
We emphasize that we have checked the cases also for other observables for a range of $\beta_{0}$-values ($\beta_0 \in [0.05, 0.3])$ and all such results show similar features as shown in  Fig.~\ref{fig:BCH_compare}(a).

Figure~\ref{fig:BCH_compare}(b) compares the coefficients of different terms computed from the BCH expansion to that obtained via our machine learning-assisted Hamiltonian reconstruction algorithm. 
The matching is quite good for most (particularly for the leading) coefficients. The slightly worse performance of our reconstruction scheme in Fig.~\ref{fig:BCH_compare}(a), compared to second-order BCH, is explained as a consequence of finite size ($N=12$) effects, causing fluctuations on the prethermal plateau and other deviations from a strictly thermal state, affecting the reconstruction input data.

%
%

\begin{figure}[t!]
\includegraphics[width=\columnwidth]{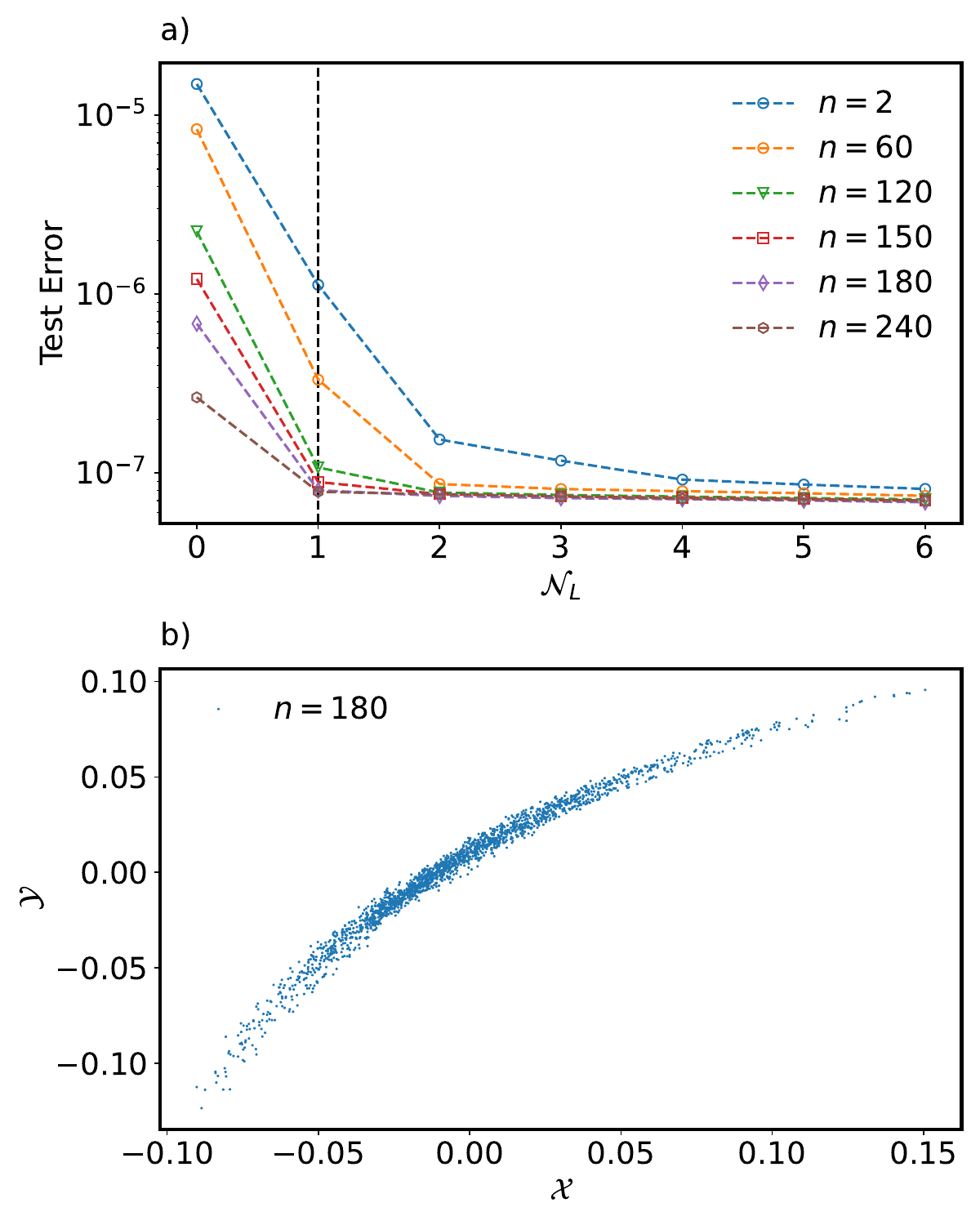}
\caption{(a) Test error as a function of the latent space dimension $\mathcal{N}_L$ (number of bottleck neurons) at different cycle number $n$ for Floquet driven systems, Eq.~\eqref{Heating_Floquet}. At smaller cycle numbers e.g., $n=2,\; 60$, the influence of initial states, prepared by applying random single-site rotations to thermal states, is still visible. For larger cycle numbers, the effectively one-dimensional latent representation signals that the system is in a thermal state throughout the entire heating regime.  (b)~Latent representation of dataset consisting of measurement at $n=180$ in the heating regime. The $x$- and $y$-axis show values of the first and the second neuron in the bottleneck layer with $\mathcal{N}_{L}=2$ neurons, giving a one-dimensional representation of data elements from different states.
Parameters: $N=18, \epsilon=0.08, J=1.0, h_x=0.9045, h_z=0.809,  T=12.7263, \beta_{0}\in[0.05, 0.45].$  
}
\label{fig:tsne_main}
\end{figure}

\begin{figure*}[t!]
\includegraphics[width=\textwidth]
{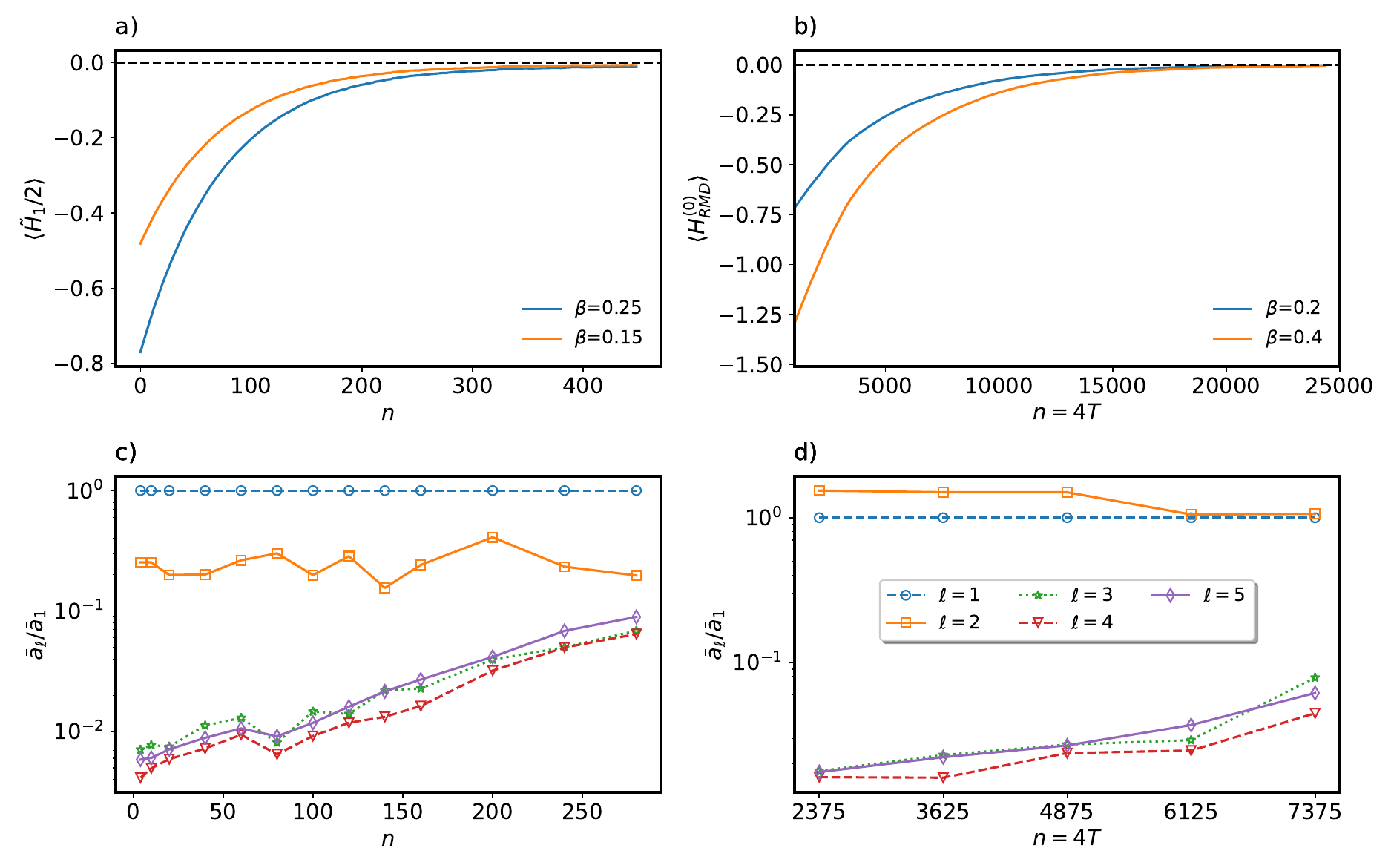}
\caption{
The considered (a) Floquet protocol, Eq.~\eqref{Heating_Floquet}, with relevant parameters $N=18, \epsilon =0.08$ and (b) random multipolar driving with $N=16, T=1/25$, Eq.~\eqref{RMDHam}, exhibit heating to an infinite-temperature steady state at easily accessible times. 
Reconstruction of the instantaneous quasistatic effective Hamiltonian, $H_{\text{eff}}(n)$, responsible for locally thermal time-dependent measurements in the heating regime, reveals a growth of averaged normalized weight at less local terms with support $\ell=3-5$ as a function of cycle number $n$, both, for (c) Floquet and (d) random multipolar driving.} 
\label{fig:Heating}
\end{figure*}

{\it Heating regime. }
We now turn our attention to the post-prethermal heating regime; we use another Floquet protocol, where the length of the prethermal plateau and the rate at which the system heats up are more easily tunable. To this end, we follow Ref.~\cite{fleckenstein21} and consider the following protocol 
\begin{align}
&U = \exp
\left(-i\tilde{H}_1T/4\right) \, \exp\left(-i \tilde{H}_2\right) \, \exp\left(-i\tilde{H}_1T/4\right), \notag \\
&\tilde{H}_1 = \sum_{i=1}^{N} J \sz_{i}\sz_{i+1} + h_z \sz_{i} + h_x \sx_{i}, \
\tilde{H}_2 = \epsilon \sum_{i=1}^{N} \sx_{i}
\label{Heating_Floquet}
\end{align} 
where $U$ is the unitary matrix for one drive cycle, which acts repeatedly on the system during the dynamics; $T$ denotes the time period of the drive.
As argued in Ref.~\cite{fleckenstein21}, the heating rate is regulated by $\epsilon$; the larger the $\epsilon$, the faster the system heats up. Here we choose $\epsilon=0.08$; Fig.~\ref{fig:Heating}(a), showing $\ave{\tilde{H}_1/2}$ at different cycle numbers $n$, confirms that heating towards the infinite temperature steady state is observed at easily accessible drive cycle numbers. We note that the period $T$ is set to $T=2(\pi k+ \epsilon)$, with $k=2$.

It has already been argued in Refs.~\cite{fleckenstein21, AnatoliPRB2021} that the system remains locally describable with a Gibbs ensemble in the whole heating regime. Here, we provide an independent check using the autoencoder analysis; it reveals that the latent representation is one-dimensional and thus is parametrized by a single parameter -- the temperature -- throughout the whole heating regime.
To confirm this, we take as initial states the thermal states with respect to the zeroth order (in $\epsilon$) Floquet Hamiltonian $\tilde{H}_1/2$ with $\beta_0 \in [0.05,0.45]$; in addition, we perform weak, random, translationally invariant single-site rotations so that the initial states manifold is no longer one-dimensional. Then we evolve such states with Floquet time evolution and probe their intrinsic dimension at chosen times.
Namely, if we considered initial states from the thermal manifold and mapped those with a time-evolution deterministic operator to states at a fixed time, from the perspective of an autoencoder, dataset remains on a one-dimensional manifold since mapping due to time propagation can be absorbed in the encoder/decoder part of the network. Therefore, to test the real, generic intrinsic dimension of measurement in the heating regime, we must start from a manifold of initial states with higher dimension that does not limit what we are probing.  
In Fig.~\ref{fig:tsne_main}(a) we show the `test error', Eq.\eqref{eq:reconstruction_loss},
as a function of the latent space dimension $\mathcal{N}_L$ (i.e., number of neurons in the bottleneck). After a few cycle numbers $n$, the `test error' drops at one latent neuron and flattens upon adding any further latent neurons, signaling that the latent representation is one dimensional. This is confirmed also if we look directly at the latent space representation of the dataset at a fixed time, cf.~Fig.~\ref{fig:tsne_main}(b).

Next, we attempt to reconstruct the effective quasistatic Hamiltonian $H_{\text{eff}}(n)$, responsible for thermal measurements in the heating regime at different cycle numbers $n$, $\ave{O(\vec\alpha)}_n = \ave{\psi(nT)|O(\vec\alpha)|\psi(nT)}\equiv \tr[O(\vec\alpha) e^{-H_{\text{eff}}(n)}/Z]$.
This is challenging, yet interesting, because the inverse frequency expansions do not converge in the heating regime.
Note that for the reconstruction of $H_{\text{eff}}(n)$, we resort to the iterative scheme
described in Sec. \ref{secII}, first selecting the leading 60 contributing
terms with maximal support $\mathcal{S}_{0}$ = 4 and then performing one iteration to increase the maximal support to $S_1$ = 5, including 160 terms in total.
The average weight of all terms in the reconstructed $H_{\text{ML}}(n)$ with support $\ell$, $\bar{a}_\ell$, relative to $\bar{a}_1$, are shown in Fig.~\ref{fig:Heating}(c) plotted as a function of cycle number $n$.
The relative growth of the weights of the less local terms with increasing $n$ indicates that $H_{\text{eff}}(n)$ becomes less and less local as the system absorbs energy from the drive. On the other hand, the Floquet Hamiltonian, defined from the unitary for one period evolution, is a time-independent object.
Our interpretation is that $H_{\text{eff}}(n)$ does not necessarily capture the stroboscopic dynamics in the 
system, but captures time-dependent thermal measurements. One way of motivating the growing support of $H_{\text{eff}}(n)$ is in terms of operator growth \cite{Ehud2019PRX}. 
Let us say we initialized the system in a thermal state with respect to $\tilde{H}_1$,  $H_{\text{eff}}(0)= \tilde{H}_1$. Time evolution of the corresponding thermal ensemble would be given by
$ U^n e^{-\tilde{H}_1} (U^\dagger)^n = e^{-U^n \tilde{H}_1 (U^\dagger)^n} = e^{-H_{\text{eff}}(n)}$, with $U$ defined in Eq.~\eqref{Heating_Floquet}. For parametrically small $\epsilon$, this should naturally lead to the parametrically regulated growth of the  $H_{\text{eff}}(n)$ operator. Our $H_{\text{ML}}(n)$ reconstruction, shown in Fig.~\ref{fig:Heating}(c), is consistent with an exponential growth with $n$.


The growing support of $H_{\text{eff}}(n)$ renders the reconstruction rather difficult, since one needs to take into consideration the contributions from terms with ever-growing support $\ell$.

\subsection{Aperiodic random multipolar drives}

Periodically driven systems obey Floquet's theorem, which guarantees the existence of a static generator of stroboscopic dynamics. Away from the strictly periodic limit, this additional structure of the time-ordered propagator is lost, and the identification of an effective Hamiltonian becomes far less obvious.

It was recently found that \textit{random multipolar driven} (RMD) systems, which are subject to an aperiodic energy-nonconserving drive, display similar prethermal properties~\cite{RoderichPRL21}:
such systems exhibit a prethermal regime (albeit with a weaker than exponential dependence of the duration of the prethermal plateau on drive frequency), followed by an eventual relaxation to an infinite temperature state. This behavior was proven mathematically by constructing an effective static Hamiltonian for the prethermal plateau using a generalization of the BCH expansion; like in Floquet systems, the expansion was shown to diverge in the heating regime \cite{KUWAHARA201696, MoriPRL2021}. To the best of our knowledge, it is unclear whether an effective Hamiltonian exists, that describes the heating dynamics of RMD systems beyond the prethermal plateau, nor what its locality properties may be. This makes RMD systems an excellent candidate to test the capabilities of our algorithm in so far uncharted territory. 

To investigate further the phenomenon of heating beyond Floquet, we consider many-body systems subject to a RMD drive. 
The protocol involves applying a random sequence of unitaries $U_{\pm}$, generated from two Hamiltonians $H_{\pm}$, each  acting for a duration $T$. The level of multipolar correlation incorporated in the drive (via the structure of the random sequence of $U_{\pm}$) is characterised by an integer $m$. For instance, $m=0$ corresponds to the case where random sequence of $U_{\pm}$ is considered, while $m=1$ represents the case for a random sequence of \textit{dipoles} having form $U_{+}U_{-}$ or $U_{-}U_{+}$. Following the same logic, $m=2$ describes a random sequence of \textit{quadrupoles} having form $U_{-}U_{+}U_{+}U_{-}$ or $U_{+}U_{-}U_{-}U_{+}$. It was demonstrated in \cite{RoderichPRL21} that RMDs, characterized by a finite value of $m$, exhibit a prethermal regime with lifetime scaling as $(1/T)^{2m+1}$. For $m \rightarrow \infty$, the limit itself corresponds to the Thue-Morse sequence constructed out of $U_{\pm}$, leads to a sub-exponential (but faster than polynomial) in $1/T$ long-lived prethermal regime.

We focus on the following Hamiltonian generators
\begin{eqnarray}
H_{\pm}=\sum_{i}J_x \sigma_{i}^{x}\sigma_{i+1}^{x}+J_z \sigma_{i}^{z}\sigma_{i+1}^{z} + B_{\pm}\sigma_{i}^{x}+
B_z \sigma_{i}^{z}
\label{RMDHam}
\end{eqnarray}
where $B_{\pm}=B_{0}\pm B_{x}$. We construct the basic unitary operators $U_{\pm}$ from $H_{\pm}$ as $U_{\pm}=\exp(-iH_{\pm}T)$. 

Figure~\ref{fig:Heating}(b) shows the behaviour of the the mean energy $ H_{\text{RMD}}^{(0)}= (H_{+}+ H_{-})/2$ for the system initialized in a thermal state with respect to $H_{\text{RMD}}^{(0)}$. As we see, $\langle H_{\text{RMD}}^{(0)} \rangle$ eventually reaches zero, indicating heating of the system to an infinite temperature state at late times. In much the same way as in the Floquet driving case, we can verify that the system remains locally thermal throughout the whole heating regime. To the best of our knowledge, this was not reported before. 

In analogy to the Floquet case, we reconstruct the effective quasistatic $H_{\text{eff}}(n)$ that captures the behavior of  local stroboscopic thermal expectation values in the heating regime. We use the iterative scheme
described in Sec. \ref{secII}, first selecting the leading 40 contributing
terms with maximal support $\mathcal{S}_{0}=4$ and then performing one further iteration to increase the maximal support to $\mathcal{S}_1=5$, including 120 terms in total.
In Fig.~\ref{fig:Heating}(d) we show the average weights $\overline{a}_\ell$ (relative to $\overline{a}_1$) of terms with support $\ell$ for
our reconstructed $H_{\text{ML}}(n)$. 

The above analysis shows that our algorithm is capable of identifying effective Hamiltonians in new setups where no analytical theory is yet available. Moreover, our results suggest the existence of simplified descriptions for the heating dynamics beyond prethermal times, where BCH-like expansions break down. Thus, the variational character of the Hamiltonian reconstruction method we propose can prove useful both in experimental nonequilibrium setups (e.g, to analyze dynamical decoupling sequences~\cite{viola1998dynamical,viola1999dynamical,bylander2011noise}, or in the digital simulation of quantum dynamics~\cite{heyl2019quantum,zhao2022making,zhao2023adaptive}), and to facilitate the theoretical analysis of nonequilibrium drives beyond Floquet systems.

\section{Conclusions and Future Outlook}
\label{conclude}

We propose a new algorithm for the Hamiltonian reconstruction from the  measurements of local operators. Reconstruction is possible under the assumption that we have access to the local measurements in different thermal states. We use autoencoder pre-processing of the data to (i) verify that the  measurements are thermal and (ii) single out the Hamiltonian candidate terms in order to reduce the reconstruction complexity. The subsequent reconstruction is accurate for local Hamiltonians as long as the dataset contains measurement of all Hamiltonian terms (of course, without knowing, which of the measured operators are the Hamiltonian terms), 
while it is only approximate for long-range interacting Hamiltonians. 

Beyond benchmarking, we apply Hamiltonian reconstruction to Floquet and random multipolar driven systems. 
For the former, we first reconstruct the effective Hamiltonian, responsible for the thermal local measurements in the prethermal plateau, and compare that to its BCH approximation. Using an autoencoder, we also confirm that the system remains locally thermal throughout the whole heating regime, and reconstruct the effective quasistatic Hamiltonian, corresponding to the Gibbs ensemble associated with the measurements at different times. We interpreted this effective quasistatic Hamiltonian's increasing support via operator growth that is parametrically regulated by the heating rate. Similar behaviour was observed also for the heating regime of the random multipolar driving. 
Despite the lack of an analytical theory, we confirmed that for the 
random multipolar driving, the heating process occurs via thermal states with growing temperature and increasingly non-local effective quasistatic Hamiltonian.


We envision our Hamiltonian reconstruction technique as a strong tool for the verification of quantum simulators that strive for the realization of exotic Hamiltonians. In our analysis, we focused on Floquet engineering, one of the most popular nonequilibrium techniques to achieve this goal, for which, however, Hamiltonians typically come with additional unwanted terms. Using autoencoder analysis to select the Hamiltonian candidate terms, we demonstrate how effective Hamiltonians can be reconstructed directly from the measurements in the prethermal plateau or the heating regime, without relying on any assumption on the Hamiltonian terms or the high-frequency expansion. 
Our operator-based Hamiltonian reconstruction is directly suitable for the verification of Hamiltonians in Rydberg atoms or trapped ion setups. 
However, our approach would have to be extended to work with partial information stored in state snapshots in cold atoms experiments.
Hamiltonian reconstruction is relevant also for condensed matter settings and quantum optics, where extracting the minimal model, responsible for the observed physics, is notoriously hard.
Similarly, due to the limited types of measurement that are accessible in these systems, the envisioned desirable generalizations of our approach call for further studies.

Another non-trivial example where our scheme could be applied, includes entanglement Hamiltonian reconstruction, at least in the intermediate to late-time regime of excited systems, where entanglement Hamiltonian approaches the actual Hamiltonian \cite{Dalmonte2022}. In future studies, it would also be desirable to develop similar strategies to reconstruct other conservation laws of integrable models that generically relax to the states that are locally describable by the generalized Gibbs ensembles.  

\vspace{4mm}
\section{acknowledgements}
It is our pleasure to acknowledge Wojciech De Roeck and Lev Vidmar for stimulating discussions. Z.L. and S.N. acknowledge the support by the QuantERA grant
QuSiED, QuantERA grant T-NiSQ, J1-2463 project, and P1-0044 program of the
Slovenian Research Agency. M.S.~was supported through the Helmholtz Initiative and Networking Fund, grant no.~VH-NG-1711. All the computations were performed at the supercomputer Vega at the Institute
of Information Science (IZUM) in Maribor, Slovenia.


%

\appendix

\section{Details of the autoencoder}\label{App:Autoencodes}
Our autoencoder consists of feed-forward neural network with the bottleneck structure. A feed-forward neural network is made up of layers, corresponding to a sequence of alternating affine-linear and non-linear transformations. The activation $a_{j}^{(l)}$ of the $l-$th layer is related to that of the previous layer $a_{j}^{(l-1)}$ as
\begin{equation}
a_{j}^{l}=\sigma\left(\sum_{k}W_{jk}a_{k}^{(l-1)}+b_{j} \right)
\label{aebasic}
\end{equation}
where $W_{jk}$ and $b_{j}$ are variational parameters and $\sigma$ denotes the fixed non-linear activate function. For all the results produced in this paper, we fix $\sigma = \tanh$. Our network consists of two encoder layers between the input and the latent space, as well as two decoder layers between the latent space and the output. Each layer has 400 neurons. 
We used the Adam optimizer for training the network and learning rate $r=0.0001$. 80\% of the data serves as the training set, while rest are used as test set.

\section{Preparing the initial thermal state}
\label{Typicality}
To prepare the initial thermal state for $N>12$, we follow the prescription detailed in Ref.~\cite{fleckenstein21}. The procedure relies on the principle of quantum typicality~\cite{BartschPRL2009, RiemannPRE2018, RiemannPRL2019, RobinPRB2019}.
Typicality states that the expectation value of an operator $\mathcal {A}$, defined on a Hilbert space $\mathcal {H}$, can be approximated as 
\begin{equation}
\frac{1}{\text{dim}\mathcal{H}}\tr[A]\approx \frac{1}{N}\sum_{n=1}^{N}\langle r_{n}|A|r_{n}\rangle
\label{eq:Typicality}
\end{equation} 
where $|r_n\rangle$ are Haar-random states. Note that the approximation becomes exact as $N\rightarrow \infty$. Therefore, thermal expectation value of observable $A$, with respect to some Hamiltonian $H$ at inverse temperature $\beta$, is given as
\begin{equation}
\langle A \rangle_{\beta} \approx \frac{\frac{\text{dim}(\mathcal{H})}{N}\sum_{n=1}^{N}\langle r_n|e^{-\beta H/2}Ae^{-\beta H/2}|r_{n}\rangle}{\frac{\text{dim}(\mathcal{H})}{N}\sum_{n=1}^{N}\langle r_n|e^{-\beta H/2}e^{-\beta H/2}|r_n\rangle}
\end{equation}
If we interpret the RHS of the above equation as an ensemble average, then the thermal density matrix can be approximated as
\begin{align}
\rho_{\beta} &\approx \frac{1}{Z_{\beta}}\frac{\text{dim}(\mathcal{H})}{N}\sum_{n=1}^{N}e^{-\beta H/2}|r_n\rangle \langle r_n|e^{-\beta H/2} \\ \nonumber
Z_{\beta} &\approx \frac{\text{dim}(\mathcal{H})}{N}\sum_{n}^{N}\langle r_n|e^{-\beta H/2}e^{-\beta H/2}|r_n\rangle
\end{align}
To time evolve the thermal density matrix, we just need to evolve a set of pure states defined as $|\psi_{n}\rangle = e^{-\beta H/2}$ and then compute
\begin{eqnarray}
\rho_{\beta}(t) \approx \frac{1}{Z_{\beta}}\frac{\text{dim}(\mathcal{H})}{N}\sum_{n=1}^{N}U(t)|\psi_{n}\rangle \langle \psi_{n}|[U(t)]^{\dagger}
\end{eqnarray}
where $U(t)$ is the evolution operator.
\vspace{4mm}
\section{BCH Expansion}
\label{BCH}
First of all, we note that for that square pulse Floquet protocol used in the main text, Floquet-Magnus expansion is identical to the BCH expansion. Mathematically, BCH formula is the solution $Z$ to the equation 
\begin{eqnarray}
    e^{X}e^{Y} = e^{Z}
\end{eqnarray}
for possibly noncommutative $X$ and $Y$ in the Lie algebra of a Lie group. BCH formula essentially yields an expression for $Z$ in terms of a formal series (not necessarily convergent) in $X$, $Y$, and iterated commutators thereof. The first few terms of the series are:
\begin{eqnarray}
\label{BCH1}
Z = X + Y + \frac{1}{2}[X,Y] + \frac{1}{12}([X,[X,Y]]-[Y,[X,Y]])+......\nonumber \\
\end{eqnarray}
To make connection with our problem, we notice that the Floquet Hamiltonian $H_{F}$ is defined via the relation 
\begin{eqnarray}
\label{BCH2}
    e^{-iH_1T/2}e^{-iH_2T/2} = e^{-iH_{F}T}
\end{eqnarray}
It is quite clear that Eq.~\eqref{BCH2} is same as Eq.~\eqref{BCH1} with $X = -iH_1T/2$ and $Y=-iH_2T/2$, where $H_1$ and $H_2$ are described by Eq.~\eqref{Pretherm_protocol_1} and Eq.\eqref{Pretherm_protocol_2} respectively, in the main text. Now, resorting to Eq.~\eqref{BCH1}, one can approximate the Floquet Hamiltonian $H_{F}$ in different orders of $T$, such that $H_{F}=\sum_{n}H_{F}^{(n)}$, where $H_{F}^{(n)}$ is proportional to $T^{n}$. We perform such calculation up to order 2 and the results are charted down below:

\begin{align}
    H_{F}^{(0)} = & \frac{1}{2}(H_1+H_2) \\
    H_{F}^{(1)} = &\frac{\gamma T}{4}\sum_{i}\sigma_{i}^{y}\sigma_{i+1}^{z} + \sigma_{i}^{z}\sigma_{i+1}^{y} + h_z\sigma_{i}^{y} \\ 
    H_{F}^{(2)} = &-\frac{T^2}{24}\gamma^{2}h_z\sum_{i}\sigma_{i}^{z} - \frac{T^2}{24}\gamma h_z^{2}\sum_{i}\sigma_{i}^{x} \\ \nonumber
    &-\frac{T^{2}}{12}\gamma^{2}\sum_{i}(\sigma_{i}^{z}\sigma_{i+1}^{z}-\sigma_{i}^{y}\sigma_{i+1}^{y}) \\ \nonumber
    &-\frac{T^2}{24}\gamma h_z \sum_{i}(\sigma_{i}^{x}\sigma_{i+1}^{z} +\sigma_{i}^{z}\sigma_{i+1}^{x})\\ \nonumber
    &+\frac{T^2}{12}\gamma h_x\sum_{i}(\sigma_{i}^{z}\sigma_{i+1}^{z} -\sigma_{i}^{y}\sigma_{i+1}^{y}) \\ \nonumber
    &-\frac{T^2}{24}\gamma h_z \sum_{i}(\sigma_{i}^{z}\sigma_{i+1}^{x} + \sigma_{i}^{x}\sigma_{i+1}^{z})\\
    \nonumber
    &-\frac{T^2}{12}\gamma\sum_{i}(\sigma_{i}^{z}\sigma_{i+1}^{z}\sigma_{i+2}^{z} + \sigma_{i}^{x})
\end{align}


\end{document}